\documentstyle[aps,preprint]{revtex}
\tighten

\begin{document}
\title{Sum rule for the twist four  
longitudinal structure function}
\author{{\bf A. Harindranath$^{a,b}$, Rajen Kundu$^{a}$,
 Asmita Mukherjee$^{a}$, and James P. Vary$^{b,c}$} \\
	$^a$Saha Institute of Nuclear Physics, 1/AF Bidhan Nagar, 
		Calcutta, 700064 India \\
	$^b$International Institute of Theoretical and Applied Physics, Iowa
State University, \\
123 Office and Lab Link, Ames, IA 50011, U.S.A. \\
        $^c$Department of Physics and Astronomy, Iowa State University, 
Ames, IA 50011, U.S.A.}
\date{October 2, 1997}
\maketitle
\begin{abstract}
We investigate the twist four longitudinal structure function $F_{L}^{\tau=4}$ of 
deep inelastic scattering and show that the integral of ${F_{L}^{\tau=4} 
\over x}$ is related to the expectation value of the fermionic part of 
the {\it light-front Hamiltonian density} at fixed momentum transfer.
We show that the new relation, in addition to providing physical intuition on 
$F_{L}^{\tau=4}$, relates the quadratic divergences of $F_{L}^{\tau=4}$ to 
the quark mass correction in light-front QCD and hence provides a new pathway 
for the renormalization of the corresponding twist four operator. The mixing of 
quark and gluon operators in QCD naturally leads to a twist four 
longitudinal gluon structure function and to a new sum rule $ \int dx 
{F_L \over x }= 4 {M^2 \over Q^2}$, which is the first sum rule obtained for 
a twist four observable. The validity of the sum rule in a non-perturbative 
context is explicitly verified in two-dimensional QCD.
    
\end{abstract}
\vskip .2in
\centerline{PACS: 11.10.Ef, 11.10.Gh, 11.55.Hx, 13.40.-f}
\vskip .2in
\noindent {\bf 1. Introduction}
\vskip .2in
Measurements \cite{exp} of $R={\sigma_L \over \sigma_T}$ show \cite{mira} that
dynamical higher twist effects play an important role in nucleon
structure experiments in the SLAC kinematic range.
Indeed, twist four ($``$power suppressed") contributions to the deep 
inelastic structure functions contain non-trivial, non-perturbative
information on the structure of hadrons \cite{pol}. 
So far, however, a clear and simple physical picture of twist
four contributions is lacking. In addition, 
power counting indicates that in the bare theory, twist four-matrix
elements will be afflicted with quadratic divergences \cite{ji}. 
Understanding the
nature of these divergences and how to remove them (renormalization) is
essential in order to compare with the experimental data.

At twist two level, light-front analysis of deep
inelastic scattering  provides an intuitive physical picture of various
structure functions. Recently, the extraction of the unpolarized 
twist two structure function $F_2$ from the transverse component of the bilocal
matrix element\cite{hari97a} and the physical picture of the transverse
polarized structure function $g_T$ and associated issues \cite{wz96,hari97b}
have been analyzed in a light-front field theory approach \cite{hari96}.
The method uses Fock space expansion for the Hamiltonian 
which is physically intuitive and is straightforward to
calculate.

To resolve the outstanding issues mentioned above,
we use this same framework
to evaluate the twist four contribution to the longitudinal structure
function. The resulting relationship with the fermionic part of the
light-front Hamiltonian density provides a convenient pathway to
renormalization and leads to a definition of a twist four longitudinal
gluon structure
function. We then find a sum rule, the first one at the twist four level.

The BJL limit, together with light-front current 
algebra\cite{jac} in $A^+=0$ gauge, the tools used in the pre-QCD era, lead
to the 
twist four part of the fermionic contribution to the longitudinal 
structure function
\begin{eqnarray}
F^{\tau=4}_{L(f)}(x) = i { 1 \over Q^2} {(xP^+)^2 \over \pi} \int dy^- e^{- {i \over 2}P^+
y^- x}  \langle P \mid \overline{\cal J}^- (y \mid 0) \mid P \rangle -
4 {(P^\perp)^2 \over Q^2} x F_{2(f)}(x) \label{flBJL1}
\end{eqnarray}
where $f$ represents a quark ($q$) or anti-quark (${\bar q}$) or both
depending on the target $ \mid P \rangle$, and
where the twist two contribution to the $F_{2(f)}$ structure function 
\begin{eqnarray}
{F_{2(f)} (x) \over x} = i { 1 \over 4 \pi }  \int dy^-  e^{- {i \over 2}P^+
y^- x}  \langle P \mid \overline{\cal J}^+(y \mid 0) \mid P \rangle .
\label{f21}  
\end{eqnarray}
The bilocal current operator,  
\begin{eqnarray}
\overline{\cal J}^\mu(y \mid 0) = { 1 \over 2 i} \left [ \overline{\psi} (y)
\gamma^\mu \psi(0) - \overline{\psi}(0) \gamma^\mu \psi(y) \right ].
\label{bcodef}
\end{eqnarray}
Note that in the case of quark (anti-quark)
contributions, the second (first) term
in the expression for the bilocal current in Eq. (\ref{bcodef}) vanishes.
We have,
\begin{eqnarray}
F^{\tau=4}_{L(q)}(x) = {\cal M}_1 + 
{\cal M}_2,  \label{fld} 
\end{eqnarray}
with
\begin{eqnarray}
{\cal M}_1~=~{ 1 \over Q^2}~ {x^2 (P^+)^2 \over 2 \pi}~ \int dy^- ~
e^{-{ i \over
2}P^+y^-x}~\langle P \mid \overline{\psi}(y^-) \gamma^- \psi(0) \mid P \rangle,
\label{flf1}
\end{eqnarray}
and
\begin{eqnarray}
{\cal M}_2=-{(P^\perp)^2  \over (P^+)^2}
{ 1 \over Q^2} {x^2 (P^+)^2 \over 2 \pi}~ 
\int dy^- ~e^{-{ i \over
2}P^+y^-x}~\langle P \mid \overline{\psi}(y^-) \gamma^+ \psi(0) \mid P \rangle.
\label{flf2}
\end{eqnarray}
Using 
 $\overline{\psi}(y^-) \gamma^- \psi(0) = 2 {\psi^{-}}^\dagger(y^-)
\psi^-(0)$, where 
\begin{eqnarray}
 \psi^-(z) = { 1 \over 4 i} \int dy^- \epsilon(z^- - y^-) \Big
[ \alpha^\perp . (i \partial^\perp + g A^\perp)+ 
\gamma^0 m \Big ] \psi^+(y^-),
\end{eqnarray}
and
$\epsilon(x^-) = - { i \over \pi} P \int {d \omega \over \omega} e^{{i \over
2 } \omega x^-}$,
we arrive at
\begin{eqnarray}
{\cal M}_1 && = { 1 \over \pi Q^2 } \int dy^- e^{-{ i \over 2}P^+ y^-x} \langle P
\mid {\psi^{+}}^\dagger(y^-) \nonumber \\
&&~~~~~~~ \Big [ \alpha^\perp. \big  [ i \partial^\perp + g A^\perp(y)
\big ] + \gamma^0 m  \Big ] \Big [ 
\alpha^\perp . \big [ i \partial^\perp + g A^\perp(0) \big ] + 
\gamma^0 m \Big ] \psi^+(0)
\mid P \rangle, \label{fl1} 
\end{eqnarray}
and 
\begin{eqnarray}
{\cal M}_2 &=&  - 2 {(P^\perp)^2  \over Q^2} { 1 \over 2 \pi} x^2 
\int dy^- e^{-
{ i \over 2} P^+y^-x} \langle  P \mid {\psi^{+}}^{\dagger}(y^-) \psi^+(0) \mid P
\rangle . 
\label{fl2}
\end{eqnarray}

Thus we have obtained an expression for the twist four part of the fermionic
component of the longitudinal structure function.
After the establishment of QCD as the underlying theory of strong
interactions, 
the twist four part of the quark contributions to 
the longitudinal structure function has been given in the
limits of vanishing target transverse momentum and 
massless quark and in the $A^+=0$ gauge using the QCD 
factorization method \cite{efp,qiu}
\begin{eqnarray}
 F^{\tau=4}_{L(q)} (x)&&= { 1 \over \pi Q^2 } \int dy^- e^{-{ i \over 2}P^+ y^-x} \langle P
\mid {\psi^{+}}^\dagger(y^-) \nonumber \\
&&~~~~~~~ \big [ i \partial^\perp + g A^\perp(y) \big ].\alpha^\perp
\alpha^\perp . \big [ i \partial^\perp + g A^\perp(0) \big ] \psi^+(0)
\mid P \rangle . \label{fact}
\end{eqnarray}

Using Eqs. (\ref{fld}), (\ref{fl1}), and (\ref{fl2}), and taking  
the limit of vanishing target transverse momentum and massless quark,
our result given in Eq. (\ref{flBJL1}) reduces to that obtained via the QCD 
factorization method (Eq. \ref{fact}).

Note that the expression for $ F_{L(q)}^{\tau=4}$ given in Eq.
(\ref{flBJL1})
appears to violate transverse boost invariance. But, we exhibit below
with explicit
calculations in Sec. 3 that the  $P^\perp$ dependence cancels between Eqs.
(\ref{fl1}) and (\ref{fl2}) so that the full $F_{L(q)}^{\tau=4}$ 
is indeed boost invariant.  
\vskip .2in
\noindent{\bf 2. Sum rule for the twist four longitudinal
structure function}
\vskip .2in

From Eq. (\ref{f21}) it follows that $ F_{2(f)}(-x) = F_{2(f)}(x)$ and 
from Eq. (\ref{flBJL1}) we explicitly find that $
F_{L(f)}^{\tau=4}(-x) = -F^{\tau=4}_{L(f)}(x) $. 
Consider the integral
\begin{eqnarray}
\int_{- \infty}^{+ \infty}dx {F^{\tau=4}_{L(f)}(x) \over x}&& = 
2 \int_{0}^{ \infty}dx
{F^{\tau=4}_{L(f)}(x)
\over x} \nonumber \\
&&= \int_{- \infty}^{+ \infty} {dx \over x}  \Bigg [
 i { 1 \over Q^2} {(xP^+)^2 \over \pi} \int dy^- e^{- {i \over 2}P^+
y^- x}  \nonumber \\
&& ~~~\times \Big [ \langle P \mid \overline{\cal J}^- (y \mid 0) \mid P \rangle -
{(P^\perp)^2 \over (P^+)^2} \langle P \mid \overline{\cal J}^+(y \mid 0) 
\mid P \rangle \Big ] \Bigg ]
\end{eqnarray}
Interchanging the orders of $x$ and $y^-$ integrations and carrying out the
integrations explicitly, we arrive at \cite{bgj}
\begin{eqnarray}
\int_{- \infty}^{+ \infty} dx  {F_{L(f)}^{\tau=4}(x,Q^2) \over x} = 
{ 4 \over Q^2} \Big [
\langle P \mid i \overline{\psi} \gamma^- \partial^+ \psi|_{(0)}
 \mid P \rangle - {(P^\perp)^2 \over (P^+)^2} 
\langle P \mid i \overline{\psi} \gamma^+ \partial^+  \psi |_{(0)} \mid P 
\rangle \Big ].
\end{eqnarray}
Identifying $ i \overline{\psi} \gamma^- \partial^+ \psi = \theta^{+-}_q$, the
fermionic part of the light-front QCD Hamiltonian density and 
$i \overline{\psi} \gamma^+ \partial^+  \psi = \theta^{++}_q$, the fermionic
part of the light-front QCD longitudinal momentum density, 
(see Eqs. (\ref{denp}) and
(\ref{denh}) below),
we have:

\begin{eqnarray}
\int_{0}^{1} dx {F_{L(f)}^{\tau=4}(x,Q^2) \over x} = { 2 \over Q^2} \Big [
\langle P \mid \theta_q^{+-}(0) \mid P \rangle - {(P^\perp)^2 \over (P^+)^2} 
\langle P \mid \theta^{++}_q(0) \mid P \rangle \Big ], \label{flsr1}
\end{eqnarray}    
where we have used the fact that the physical structure function 
vanishes for $x >1$.

The integral of ${F^{\tau=4}_{L(f)} \over x}$ is therefore 
related to the hadron matrix
element of the (gauge invariant) fermionic part of the light-front 
{\it Hamiltonian density}.
This result
manifests the physical content and the non-perturbative nature of the 
twist-four part of the
longitudinal structure function.

The fermionic operator matrix elements appearing in Eq. (\ref{flsr1}) 
change with $Q^2$ as a result of the mixing of quark and gluon operators in
QCD under renormalization.    
Next we analyze the operator mixing and derive a new 
sum rule at the twist four level. 

The symmetric, gauge-invariant energy-momentum tensor in QCD is
given by
\begin{eqnarray}
\theta^{\mu \nu} = && { 1\over 2} \overline{\psi} i \big [ \gamma^\mu D^\nu +
\gamma^\nu D^\mu \big ] \psi 
-F^{\mu \lambda a} F^{\nu}_{~ \lambda a} + {1 \over 4} g^{\mu \nu} 
(F_{\lambda \sigma a } )^2 \nonumber \\
&& -g^{\mu \nu} \overline{\psi} \left ( i \gamma^\lambda D_\lambda -
m  \right ) \psi.
\label{emt}
\end{eqnarray}
The last term vanishes using the equation of motion.

{\it Formally,} we split the energy momentum tensor into a 
$``$fermionic" part $\theta^{\mu \nu}_{q}$ representing the first term in
Eq. (\ref{emt}) and a $``$gauge bosonic" 
part $ \theta^{\mu \nu}_g$ representing the second and third terms in Eq.
(\ref{emt}).
To be consistent with the study of
deep inelastic structure function formulated in the $A^+=0$ gauge,
we shall work in the same gauge. 
We have, for the fermionic part of the longitudinal momentum density,
\begin{eqnarray}
\theta^{++}_{q} = i \overline{\psi} \gamma^+ \partial^+ \psi. \label{denp} 
\end{eqnarray}
For the fermionic part of
the Hamiltonian density, we have, 
\begin{eqnarray}
\theta^{+-}_q = i {\psi^{+}}^\dagger \partial^- \psi^+ + g {\psi^{+}}^\dagger
A^- \psi^+ + i {\psi^{-}}^\dagger \partial^+ \psi^-.
\end{eqnarray}
Using the Dirac equation for the fermion, we find that the sum of the 
first two terms equals the third term in the above equation. Therefore, 
\begin{eqnarray}     
\theta^{+-}_q &&= i \overline{\psi} \gamma^- \partial^+ \psi =
2 i {\psi^{-}}^\dagger \partial^+ \psi^-  \label{denh} \\
&& = 2 {\psi^{+}}^\dagger \Big [ \alpha^\perp.(i \partial^\perp + g A^\perp) 
+ \gamma^0 m \Big ]
{ 1 \over i \partial^+} \Big [ \alpha^\perp . (i \partial^\perp + g A^\perp)
+ \gamma^0 m \Big ] \psi^{+} \label{thetaqf}.
\end{eqnarray}

The gauge boson part of the Hamiltonian density is given by
\begin{eqnarray}
\theta^{+-}_g && =  - F^{+ \lambda a} F^{-}_{\lambda a}+ { 1 \over 4} 
g^{+-} (F_{\lambda \sigma a})^2 =
{ 1 \over 4} \Big (\partial^+ A^{- a}\Big )^2 + 
{ 1 \over 2} F^{ij a } F^{a}_{ij}
\nonumber \\
&& = (\partial^i A_a^j)^2 + 2gf^{abc}A_a^i A_b^j \partial^i A_c^j
		  + \frac{g^2}{2}
		f^{abc} f^{ade} A_b^i A_c^j A_d^i A_e^j  \nonumber \\
	&& ~~~~~~~~~~ + 2g \partial^i A_a^i \left( \frac{1}{\partial^+}
		\right) (f^{abc} A_b^j \partial^+ A_c^j + 2 (\psi^+)^{\dagger}
		T^a \psi^+ ) \nonumber \\
	&& ~~~~~~~~~~ + g^2 \left( \frac{1}{\partial^+}
		\right) (f^{abc} A_b^i \partial^+ A_c^i + 2 (\psi^+)^{\dagger}
		T^a \psi^+ ) 
	  \left( \frac{1}{\partial^+}\right)
		(f^{ade} A_d^j \partial^+ A_e^j + 2 (\psi^+)^{\dagger} T^a
		\psi^+ )     \nonumber \\
    \label{thetagf}
\end{eqnarray}
where we have used the equation of constraint for the gauge field.     

We define the twist four longitudinal gluon structure function
\begin{eqnarray}
F_{L(g)}^{\tau=4}(x) && = { 1 \over Q^2} {x P^+ \over 2 \pi} \int dy^- ~
e^{-{i \over 2} P^+ y^- x} \nonumber \\
&& ~~~~~\Big [ \langle P \mid (-) F^{+ \lambda a}(y^-) F^-_{\lambda a}(0) + 
{ 1 \over 4} g^{+-} F^{\lambda \sigma a} (y^-) F_{\lambda \sigma a}(0) \mid
P \rangle \nonumber \\
&& ~~~~~~~ - {(P^\perp)^2 \over (P^+)^2} \langle P \mid F^{+ \lambda a}(y^-) 
F^+_{\lambda a}(0) \mid P \rangle \Big ].
\end{eqnarray}
Then we have,
\begin{eqnarray}
\int_0^1 { dx \over x} \Big [ F_{L(q)}^{\tau=4} + F_{L(g)}^{\tau=4} \Big ] =
\int_0^1 { dx \over x} F_{L}^{\tau=4} = { 2 \over Q^2} 
\Big [ \langle P \mid \theta^{+-}(0) \mid P \rangle - {(P^\perp)^2 \over
(P^+)^2} \langle P \mid \theta^{++}(0) \mid P \rangle \Big ].
\label{flsr2}
\end{eqnarray}
But,
\begin{eqnarray}
\langle P \mid \theta^{+-}(0) \mid P \rangle = 2 P^+ P^- = 2 (M^2 +
(P^\perp)^2)~~ {\rm and} ~~ \langle P \mid \theta^{++}(0) \mid P \rangle = 2
(P^+)^2,
\end{eqnarray}
where $M$ is the invariant mass of the hadron.
Thus we arrive at the new sum rule for the twist four part of the
longitudinal structure function
\begin{eqnarray}
\int_0^1 { dx \over x} F_L^{\tau=4} = 4 {M^2 \over Q^2}. \label{flsr}   
\end{eqnarray}
To our knowledge, this is the first sum rule at the twist four level of deep
inelastic scattering or for QCD in general.
\eject
\vskip .2in
\noindent {\bf 3. Implications of the sum rule}
\vskip .2in

Next, we investigate the implications of Eq. (\ref{flsr1})
for quadratic divergences in $F_{L(q)}^{\tau=4}$.
For simplicity, we select a dressed quark target 
and evaluate the structure functions 
to order $g^2$. That is, we take the state 
$ \mid P \rangle$ to be a dressed quark
consisting of bare states of a quark and a quark plus 
a gluon:
\begin{eqnarray}
\mid P, \sigma \rangle && = \phi_1 b^\dagger(P,\sigma) \mid 0 \rangle
\nonumber \\  
&& + \sum_{\sigma_1,\lambda_2} \int 
{dk_1^+ d^2k_1^\perp \over \sqrt{2 (2 \pi)^3 k_1^+}}  
\int 
{dk_2^+ d^2k_2^\perp \over \sqrt{2 (2 \pi)^3 k_2^+}}  
\sqrt{2 (2 \pi)^3 P^+} \delta^3(P-k_1-k_2) \nonumber \\
&& ~~~~~\phi_2(P,\sigma \mid k_1, \sigma_1; k_2 , \lambda_2) b^\dagger(k_1,
\sigma_1) a^\dagger(k_2, \lambda_2) \mid 0 \rangle. 
\end{eqnarray}

Without loss of generality,
 we work in the massless limits for the
dressed and bare quarks to obtain
\begin{eqnarray}
{\cal M}_1 &&= { 1 \over \pi Q^2 } \int dy^- e^{-{ i \over 2}P^+ y^-x} \langle P
\mid {\psi^{+}}^\dagger(y^-) \nonumber \\
&&~~~~~~~ \big [ i \partial^\perp + g A^\perp(y) \big ].\alpha^\perp
\alpha^\perp . \big [ i \partial^\perp + g A^\perp(0)  \big ] \psi^+(0)
\mid P \rangle, \label{fls1} 
\end{eqnarray}
and 
\begin{eqnarray}
{\cal M}_2 = - 4{ (P^\perp)^2 \over Q^2}   x F_{2(q)}(x).
\label{fls2}
\end{eqnarray}

Note that the matrix elements appearing in Eqs. (\ref{fl1}), (\ref{fl2}), 
(\ref{thetaqf}), and (\ref{thetagf})
involve products of operators. In this initial study of quadratic
divergences we treat them as normal ordered which is sufficient for our
purposes here.
This aspect of the problem
merits further investigation which we hope to undertake in the near future.
However, the terms we drop are quadratically divergent and they will affect
only the counterterm structure.

First we evaluate the contribution ${\cal M}_2$ given in 
Eq. (\ref{fls2}).
We obtain,
\begin{eqnarray}
{\cal M}_2= -4 C_f{(P^\perp)^2 \over Q^2} x^2 \Big [ \delta(1-x) + {\alpha_s \over 2
\pi} ln \Lambda^2 \big [ { 1+x^2 \over 1-x} - \delta(1-x) \int dy {1+y^2 \over 1-y}
\big ] \Big ]\label{fl2f}
\end{eqnarray}
where we have cutoff the transverse momentum integral at $\Lambda$.
Note that the result in Eq. (\ref{fl2f}) violates transverse boost invariance. 

We have, from Eq. ({\ref{fls1}),
\begin{eqnarray}
{\cal M}_1 
&& = -{ 1 \over \pi Q^2} \int dy^- e^{-{i \over 2} P^+ y^- x} \langle P \mid
{\psi^{+}}^\dagger(y^-) (\partial^\perp)^2 \psi^+(0) \mid P \rangle
\nonumber \\
&& ~~~+ g{ 1 \over \pi Q^2} \int dy^- e^{-{i \over 2} P^+ y^- x} \langle P \mid
{\psi^{+}}^\dagger(y^-) i \partial^\perp . \alpha^\perp \alpha^\perp.
A^\perp(0) \psi^+(0) \mid P \rangle \nonumber \\
&& ~~~+ g{ 1 \over \pi Q^2} \int dy^- e^{-{i \over 2} P^+ y^- x} \langle P \mid
{\psi^{+}}^\dagger(y^-) \alpha^\perp.A^\perp(y) i \partial^\perp .\alpha^\perp
\psi^+(0) \mid P \rangle \nonumber \\
&& ~~~+ g^2{ 1 \over \pi Q^2} \int dy^- e^{-{i \over 2} P^+ y^- x} \langle P \mid
{\psi^{+}}^\dagger(y^-) A^\perp(y).A^\perp(0) \mid P \rangle \\
&&\equiv {\cal M}_1^{a}+{\cal M}_1^{b}+{\cal M}_1^{c}+{\cal M}_1^{d}.
\end{eqnarray} 
Since the operators in Eq. (\ref{fls1}) are taken to be normal ordered, 
the
contribution of ${\cal M}_1^{d}$ vanishes to order $g^2$.
 
Explicit calculation leads to the diagonal Fock basis contributions  
\begin{eqnarray}
({\cal M}_1)_{diag}=
{\cal M}_1^{a} = && 4 C_f{(P^\perp)^2 \over Q^2} x^2 \Big [ \delta(1-x) + {\alpha_s
\over 2 \pi} ln \Lambda^2 \big [ {1+x^2 \over 1-x} - \delta(1-x) \int dy { 1+ y^2
\over 1-y} \big ] \Big ] \nonumber \\
&& ~~~~+  C_f{g^2 \over 2 \pi^2} {\Lambda^2 \over Q^2}{ 1+x^2 \over 1-x} .
\end{eqnarray}
The first term here explicitly cancels the term ${\cal M}_2$ given in
Eq. (\ref{fl2f}).

Off-diagonal contributions
\begin{eqnarray}
({\cal M}_1)_{nondiag}={\cal M}_1^{b}+{\cal M}_1^{c} = 
-C_f{g^2 \over \pi^2} {\Lambda^2 \over Q^2}
{1 \over 1-x}.
\end{eqnarray} 

Adding all the contributions, we have, 
\begin{eqnarray}
F^{\tau=4}_{L(q)}(x) = - C_f{ g^2 \over 2 \pi^2} 
{\Lambda^2 \over Q^2}(1+x).
\label{fltot}
\end{eqnarray}


As anticipated from power counting, we have generated quadratic
divergences for $F_L$ in the bare theory. Now we are faced with two
issues:
(a) What is the principle for adding counterterms? (b) What
determines the finite parts of the counterterms?

Further, since $F^{\tau=4}_{L(q)}$ is directly related to the 
physical longitudinal cross
section, we expect $F^{\tau=4}_{L(q)}$ to be positive definite 
(see, for example, Ref.
\cite{efp}). 
From our answers we see that we get a negative answer which is free 
from end point singularities but is quadratically divergent. 
Note that it is easy to show from our expressions in Eqs. (\ref{fl1}) and
(\ref{fl2})
for a free quark of mass $m$,
$F^{\tau=4}_{L(q)}=4 {m^2 \over Q^2} \delta(1-x) $, a
well-known result.



Now we show that the sum rule relating the integral of ${F^{\tau=4}_{L(q)}(x)
\over x}$ to the fermionic part of the light-front Hamiltonian density helps
us to understand the results we obtained for $F_L$ for a dressed quark in
the bare theory and indicates how to add counterterms to renormalize 
the theory.

A straightforward evaluation leads to
\begin{eqnarray}
\langle P \mid \theta^{+-}_q(0) \mid P \rangle_{nondiag}  = 
-C_f{ 1 \over 2}{g^2
\over \pi^2} \Lambda^2  \int_0^1
{dx \over x} { 1 \over 1-x}, \label{tfode}
\end{eqnarray}
\begin{eqnarray} 
\langle P \mid \theta^{+-}_q(0) \mid P \rangle_{diag} - 
{(P^\perp)^2 \over (P^+)^2}
\langle P \mid \theta^{++}_{q}(0) \mid P \rangle_{diag} = C_f{ 1 \over 2}{g^2
\over 2 \pi^2} \Lambda^2  \int_0^1
{dx \over x} { 1  + x^2 \over 1-x},\label{tfde}
\end{eqnarray}

Adding the diagonal and off-diagonal contributions from the fermionic part
of the Hamiltonian density, we arrive at
\begin{eqnarray}
{2 \over Q^2} \Big [ \langle P \mid \theta^{+-}_q(0) \mid P \rangle 
- {(P^\perp)^2 \over (P^+)^2}\langle P \mid \theta^{++}_q(0) \mid P \rangle  \Big ] =
- C_f { g^2 \over 2 \pi^2} 
{\Lambda^2 \over Q^2}\int_0^1 { dx \over x} (1+x).
\label{thetatot}
\end{eqnarray} 

Comparison of Eqs. (\ref{fltot}) and (\ref{thetatot}) immediately shows that
the relation given in Eq. (\ref{flsr1}) has been successfully tested to order $g^2$ in perturbative
QCD in the bare theory.


Apart from the verification of the sum rule, the dressed quark calculation
serves to illustrate the {\it physical origin} of quadratic divergences in
the corresponding twist four matrix element. The sum rule explicitly relates
the quadratic divergences in $F^{\tau=4}_{L(q)}$ to the fermion mass shift
in light-front Hamiltonian perturbation theory. The quadratic divergence is
to be removed by adding counterterms. The precise selection of counterterms
is dictated by the regularization and renormalization of the light-front QCD
Hamiltonian, the diagonalization of which results in the hadronic structure.
The choice of counterterms in the Hamiltonian, in turn, dictates the
counterterms to be added to the longitudinal structure function which
results in the prediction of theoretical twist four contribution to $F_L$. 
 
To complete the discussion of the fermion mass shift, we now consider the 
contributions from the gluonic part of the Hamiltonian
density. 

Take the off-diagonal contribution arising from 
\begin{eqnarray}
{\cal M}_3= -\langle P \mid 4 g { 1 \over \partial^+} \Big ( \partial^i
A^{ia}
\Big ) {\psi^{+}}^\dagger T^a \psi^+ \mid P \rangle.
\end{eqnarray}
A straightforward evaluation leads to 
\begin{eqnarray}
{\cal M}_3 = -C_f{g^2 \over 2 \pi^2}
\Lambda^2 \int_0^1 { dx \over x} \Big [ { 2x^2 \over (1-x)^2} + {x \over 1-x}
\Big ]. \label{tgode}
\end{eqnarray}
Thus from Eqs. (\ref{tfode}) and (\ref{tgode}) we have,
\begin{eqnarray}
\langle P \mid \Big [ \theta^{+-}_{q} (0) + \theta^{+-}_{g} (0) \Big
] \mid P
\rangle_{nondiag} = - C_f{ g^2 \over 2 \pi^2} \Lambda^2 \int_0^1 { dx \over x} {1 + x^2
\over (1-x)^2}. \label{todtot}
\end{eqnarray}

Next let us consider diagonal contributions.
From the gauge boson part of the Hamiltonian density, we have,
\begin{eqnarray}
\langle P \mid  \theta^{+-}_{g} (0)\mid P \rangle_{diag} 
 - {(P^\perp)^2 \over (P^+)^2} \langle P \mid  \theta^{++}_{g}(0)  
\mid P \rangle_{diag} 
= C_f{ g^2 \over 4 \pi^2} \Lambda^2 \int_0^1 { dx \over 1-x}{1 + x^2
\over 1-x}. \label{tgde}
\end{eqnarray}
Adding the diagonal contributions from the fermion and gauge boson parts,
i.e., Eqs. (\ref{tfde}) and (\ref{tgde}) we
arrive at
\begin{eqnarray}
\langle P \mid \theta^{+-}(0) \mid P \rangle_{diag} - {(P^\perp)^2 \over
(P^+)^2} \langle P
\mid \theta^{++}(0) \mid P \rangle_{diag} = C_f{ g^2 \over 4 \pi^2} \Lambda^2
\int_0^1 { dx \over x} {1+x^2 \over (1-x)^2}. \label{tdtot}
\end{eqnarray}
Thus the total contribution (quark and gluons) to the expectation
value of the Hamiltonian density is given by
\begin{eqnarray}
\langle P \mid  \theta^{+-} (0) \mid P \rangle - 
{(P^\perp)^2 \over (P^+)^2} \langle P \mid \theta^{++}(0) 
\mid P \rangle  = - C_f{ g^2 \over 4 \pi^2} \Lambda^2 \int_0^1 {dx \over x} { 1+x^2 \over
(1-x)^2}.
\end{eqnarray}
This result is directly related to the mass shift of the fermion to order
$g^2$ in light-front perturbation theory. We have in the limit of zero bare
mass, (see Eq. (4.10) in Ref. \cite{qcd2}),    
\begin{eqnarray}
\delta p_1^- = -{ 1 \over 2 P^+} C_f{g^2 \over 4 \pi^2} \Lambda^2 \int_0^1
{dx \over x} {1+x^2 \over (1-x)^2}.
\end{eqnarray}

Note that in the massless limit we encountered only quadratic divergences in the twist
four part of the longitudinal structure function. It is now obvious that for
a massive quark, we will also encounter
logarithmic divergences.

To test the sum rule given in Eq.
(\ref{flsr}) explicitly in a non-perturbative context we turn to 
two-dimensional QCD.
In 1+1 dimensions, in $A^+=0$ gauge, we have,
\begin{eqnarray}
\int_0^1 {dx \over x} F_{L(q)}^{\tau=4}(x) = { 2 \over Q^2} \langle P \mid 
\Big [ \theta^{+-}_q(0) + \theta^{+-}_g(0) \Big ] \mid P \rangle,
\end{eqnarray} 
with $
\theta^{+-}_q = 2 m^2 {\psi^{+}}^\dagger { 1 \over i \partial^+} \psi^+ $
and $ \theta^{+-}_g = - 4 g^2 {\psi^{+}}^\dagger T^a \psi^+ 
{1 \over (\partial^+)^2}
{\psi^{+}}^\dagger T^a \psi^+$.
In the standard one pair ($q \overline{q}$) approximation to the ground state,
with the help of the bound state equation ('t Hooft equation) obeyed by the
ground state wave function $\psi(x)$ for the meson 
\begin{eqnarray}
M^2 \psi(x) = {m^2 \over x (1-x)} \psi(x) - C_f {g^2 \over \pi} \int dy {
\psi(y) - \psi(x) \over (x-y)^2}   
\end{eqnarray}
together with the normalization condition $ \int_0^1 dx \psi^2(x)=1$, it is
easily verified that the twist four longitudinal structure function of the
meson obeys the sum rule
\begin{eqnarray}
\int_0^1 {dx \over x} F_L^{\tau=4} = { 2 \over Q^2} \langle P \mid
\theta^{+-}(0) \mid P \rangle = 4 {M^2 \over Q^2}.
\end{eqnarray}

To summarize, we have investigated the twist four contributions to the
longitudinal structure function in deep inelastic scattering
in an approach based on Fock space expansion
methods in light-front field theory. Our
motivation has been 
to gain physical intuition on the twist four part of the longitudinal
structure function and 
to understand the occurrence of quadratic divergences in
$F_{L(q)}^{\tau=4}$ and the associated renormalization issues. 
We have found that
the integral of ${F_{L(q)}^{\tau=4}(x) \over x}$ is related to 
the expectation value of
the fermionic part of the light-front Hamiltonian density. The relation,
which is tested to order $g^2$ in perturbative QCD in the bare
theory, explicitly
relates the quadratic divergences generated in $F^{\tau=4}_{L(q)}$ 
to the fermion mass shift arising from the fermionic part of the light-front
Hamiltonian density. By investigating the operator mixing
problem, we have derived a new sum rule which relates the integral of
 ${F^{\tau=4}_L(x) \over x}$ to the invariant mass of the hadron.
We have explicitly checked the validity of 
this sum rule in two dimensional QCD.

The phenomenological
consequences of the new sum rule are worth studying. 
Clearly, experiments to measure the twist four longitudinal structure
function will reveal the fraction of the hadron mass carried by 
the charged parton components.
These experiments will complement the longitudinal momentum and helicity
distribution information obtained at the twist two level. It is of interest
to investigate the feasibility of the direct measurement of the twist four
gluon structure function in high energy experiments. Recent work of Qiu,
Sterman and collaborators have shown that semi-inclusive single jet
production in deep inelastic scattering \cite{luo} and direct photon
production in hadron nucleus scattering \cite{guo} provide direct measurement 
of twist
four gluon matrix elements.

On the theoretical side we note that
some significant progress has been made recently in the bound state problem
in light-front QCD \cite{bsp}. 
In the near future, we plan to undertake a non-perturbative calculation 
(utilizing Fock space expansion and Hamiltonian
renormalization techniques) of the longitudinal structure function for a
meson-like bound state. Such a calculation will undoubtedly help to
check the validity of current phenomenological models \cite{mira} based on
simple assumptions \cite{simple} employed
in analyzing the data.  
Whether the relation between the integral of
${F_{L(q)}^{\tau=4}(x) \over x}$ and the hadron expectation value of the
fermionic part of the light-front QCD Hamiltonian density can provide
insight in to the small $x$ behavior of $F_{L(q)}^{\tau=4}$ is also worth
investigating.

One of the authors (AH) would like to thank Wei Min Zhang for useful
discussions. This work is supported in part by the U.S. Department of Energy
under Grant No. DEFG02 - 87ER40371, Division of High Energy and Nuclear
Physics.



\end{document}